\documentclass[11pt]{article}
\usepackage{osid}

\usepackage[english]{babel}
\usepackage{textcomp,xcolor}
\usepackage{dcolumn}
\usepackage{graphicx}
\graphicspath{{Figures/}}

\usepackage{amsmath, bbm}
\usepackage{latexsym}
\usepackage{amsfonts}   
\usepackage{amssymb}
\usepackage{array}      
\usepackage{epsfig}
\usepackage{txfonts}
\usepackage{xcolor,braket}
\usepackage[colorlinks=true,linkcolor=blue,urlcolor=blue,citecolor=blue]{hyperref}
\usepackage[normalem]{ulem}
\usepackage{soul}
\usepackage{dsfont}

\usepackage{xfrac}  
\usepackage{relsize}

\usepackage{amsthm}
\usepackage{mathrsfs, mathtools, amsmath}

\usepackage[normalem]{ulem}
\usepackage{cancel}
\usepackage{enumitem}

\usepackage{xspace}

\usepackage[style=phys]{biblatex}
\bibliography{biblio}

\newcommand{\ketbra}[1]{\ket{#1}\bra{#1}}
\newcommand{\id}{\mathbbm{1}}
\newcommand{\abs}[1]{\lvert #1 \rvert}
\newcommand{\norm}[1]{\lVert #1 \rVert}
\newcommand{\tr}{\operatorname{tr}}
\newcommand{\RO}{$\Psi$-RO\xspace}
\newcommand{\ro}{\RO}
\newcommand{\schro}{Schr\"odinger\xspace}

\title{A Stochastic \schro Equation for the Generalized Rate Operator Unravelings}
\author{Federico Settimo\\\it
Department of Physics and Astronomy, University of Turku, FI-20014 Turun yliopisto, Finland\\
e-mail: fesett@utu.fi}

\begin{document}

\maketitle

\begin{abstract}
    Stochastic unravelings are a widely used tool to solve open quantum system dynamics, in which the exact solution is obtained via an average over a stochastic process on the set of pure quantum states.
    Recently, the generalized rate operator unraveling formalism was derived, allowing not only for an engineering of the stochastic realizations, but also to unravel without reverse jumps even for some dynamics in which P-divisibility is violated, thus hugely improving the simulation efficiency.
    This is possible because the unraveling depend on an arbitrary non-linear transformation which can incorporate the memory effects.
    In this work, a stochastic Schr\"odinger equation for this formalism is derived, both for cases with and without reverse jumps.
    It is also shown that a failure of this method can be used to witness master equations leading unphysical time evolutions, independently on the particular non-linear transformation considered.
\end{abstract}

% ------------------------------------ Intro ------------------------------------
\section{Introduction}
\label{sec:intro}
The dynamical evolution of open quantum systems can be described by a completely positive (CP) trace preserving map $\Lambda_t:\rho(0)\mapsto\rho(t)=\Lambda_t[\rho(0)]$ \cite{Breuer-oqs, Vacchini-OQS}.
Such map is the solution of the master equation (ME) $d\rho/dt = \mathcal L_t[\rho]$, where the generator $\mathcal L_t = \dot\Lambda_t\Lambda_t^{-1}$ can be written as \cite{Gorini1976, Lindblad1976}
\begin{equation}
    \label{eq:ME}
    \mathcal L_t[\rho] = -i[H(t),\rho]+\sum_\alpha\gamma_\alpha(t) L_\alpha(t)\rho L_\alpha^\dagger(t)-\frac12\{\Gamma(t),\rho\},
\end{equation}
where $\Gamma(t) = \sum_\alpha\gamma_\alpha(t) L_\alpha^\dagger(t) L_\alpha(t)$ and both the rates $\gamma_\alpha(t)$ and the operators $H(t)$ and $L_\alpha(t)$ can depend on time.

The rates $\gamma_\alpha(t)$ can be temporarily negativity, without violating the complete positivity of the dynamical map $\Lambda_t$ \cite{Breuer-oqs, Vacchini-OQS, Chruscinski-nM-local, Chruscinski2022}.
However, positivity of the rates is equivalent to being able to decompose $\Lambda_t=\Lambda_{t,s}\Lambda_s$ with CP operators $\Lambda_{t,s}$ for all times $t\ge s\ge0$.
If this is the case, the dynamics is said to be CP-divisible.
If, instead, $\Lambda_{t,s}$ is only positive, then the dynamics is said to be P-divisible, which is equivalent to the weaker condition
\cite{Kossakowski-necessary, BLPV-colloquium}
\begin{equation}
    \label{eq:P_div}
    \sum_j\gamma_j(t)\abs{\braket{\varphi_\mu\vert L_j(t)\vert\varphi_{\mu^\prime}}}^2\ge0
\end{equation}
for all orthonormal bases $\{\varphi_\mu\}_\mu$ and for all $\mu\ne\mu^\prime$.
Violations of both P- and CP-divisibility have been connected to non-Markovianity and memory effects \cite{BLPV-colloquium, rivas-quantum-nm, BLP, BLP-PRA, RHP, Chruscinski-nM, Wißmann2015}.

Obtaining an analytical solution to the ME \eqref{eq:ME} is in general a very difficult task, and numerical methods are generally used to approximate the exact solution.
A widely used numerical tool is stochastic unravelings, in which the solution $\rho(t)$ is obtained as the average over a random process on the set of pure states
\begin{equation}
    \label{eq:avg_rho}
    \rho(t) = \int d\psi\,p(\psi,t)\ketbra\psi \approx\sum_i\frac{N_i(t)}N\ketbra{\psi_i(t)},
\end{equation}
where $p(\psi,t)$ is a probability distribution.
In numerical methods, such probability distribution is generally approximated as $p(\psi_i,t)\approx N_i(t)/N$, where $N_i(t)$ is the number of stochastic realizations in the state $\ket{\psi_i(t)}$ and $N=\sum_iN_i(t)$ is the total number of realizations.
Unravelings can be divided into two major families: the underlying process can be either diffusive \cite{Percival-qsd, Barchielli-traj, Gisin1992, Diosi-NMQSD, Yu-NMQSD-PRL, Caiaffa-W-diffusive, Luoma-diffusive-NMQJ} or it can be piecewise deterministic and interrupted by discontinuous jumps \cite{Plenio-jumps-review, Diosi-orthogonal-jumps, Dalibard-MCWF, Dum-MonteCarlo-emission, Daley-jumps-many-body, Smirne-W, Chruscinski-Quantum-RO, Settimo2024}.
Such stochastic methods can be generalized to the case in which the system and the environment are initially correlated \cite{Settimo-OPD}.

The probability distribution $p(\psi,t)$ can be generated in multiple non-equivalent ways, and different unraveling techniques can be applied or not depending on the divisibility properties of the dynamics.
For indivisible dynamics, the piecewise deterministic unravelings can be equipped with reverse jumps \cite{Piilo-NMQJ-PRL, Piilo-NMQJ-PRA}.
This method, however, makes the simulations more expensive, since the different stochastic realizations are no longer independent.
Recently, in \cite{Settimo2024}, the generalized rate operator (\ro) unraveling method was derived and shown to give independent realizations for all P-divisible and even some indivisible dynamics.
This is possible thanks to a non-linear transformation used in the definition of the jump process, which can capture the memory effects present in the dynamics.
This non-linearity allows also for a great control in the stochastic realizations, thus improving the efficiency of the simulations.
In this work, a stochastic \schro equation (SSE) for the unravelings obtained with the \ro formalism is derived, both with and without reverse jumps.
This gives a proper formalization of the technique and it is shown that the noise average solution of the SSE obeys the ME \eqref{eq:ME}.
Failure of such method can be used to witness violations of positivity of the map $\Lambda_t$, thus ruling out unphysical evolutions.
Noticeably, this criterion does not depend on the particular non-linear transformation considered, and therefore it is easily evaluated.

The rest of the paper proceeds as follows.
In Sec.~\ref{sec:RO}, the piecewise deterministic process obtained via the \ro formalism of \cite{Settimo2024} is presented, both with and without reverse jumps.
Sec.~\ref{sec:SSE} contains the main results of this work, in which the SSE is derived and it is shown that its solution obeys the ME \eqref{eq:ME} on average.
It is also shown that failure of such SSE implies unphysicality of the dynamical evolution.
In Sec.~\ref{sec:example}, an example of application of the SSE is presented, in which a non-P-divisible dynamics is unraveled  by employing the flexibility of the \ro.
Finally, in Sec.~\ref{sec:conclusion}, the conclusion of the work are presented.

% ------------------------------------ RO ------------------------------------
\section{Generalized Rate Operator unravelings}
\label{sec:RO}
The ME \eqref{eq:ME} can be divided into a jump and a driving term, that read, respectively,
\begin{equation}
    \label{eq:jump_ME}
    \mathcal J_t[\rho]\coloneqq \sum_\alpha\gamma_\alpha L_\alpha\rho L_\alpha^\dagger,\qquad
    \mathcal D_t[\rho] \coloneqq -i\left(K_t\,\rho-\rho\,K^\dagger_t\right),
\end{equation}
with the non-Hermitian Hamiltonian $K_t \coloneqq H-i \Gamma/2$ and $\mathcal L_t = \mathcal D_t+\mathcal J_t$.
The explicit dependence on time has been suppressed.
This division, however, is not unique: any transformation of the form \cite{Chruscinski-Quantum-RO}
\begin{gather}
    \mathcal J_t^\prime[\rho]\coloneqq\mathcal J_t[\rho]+\frac12\left(C_t\,\rho+\rho\,C_t^\dagger\right),\qquad
    K^\prime_t\coloneqq K_t-\frac i2 C_t,
\end{gather}
where $C_t$ is an arbitrary (eventually time-dependent) operator, leaves $\mathcal L_t$ unchanged.
The transformed operator $\mathcal J_t^\prime$ is known as the rate operator \cite{Diosi-orthogonal-jumps, Caiaffa-W-diffusive, Smirne-W}.
In \cite{Chruscinski-Quantum-RO}, this freedom was used to derive different unraveling schemes depending on the transformation $C_t$, thus allowing for engineering of the stochastic realizations.

In \cite{Settimo2024}, it was shown that, from the point of view of a single stochastic realization $\ket{\psi}$, the transformation $C_t$ can depend not only on time but also on $\psi$: $C_t\mapsto C_{\psi,t}$.
This observation leads to the  definition of the generalized rate operator (\ro)
\begin{equation}
    \label{eq:RO}
    R_\psi \coloneqq \sum_\alpha\gamma_\alpha L_\alpha\ketbra{\psi}L_\alpha^\dagger+\frac12\left(\ket\psi\bra{\Phi_\psi}+\ket{\Phi_\psi}\bra\psi\right),
\end{equation}
where $\ket\psi$ is the state of the realization and $\ket{\Phi_\psi}\coloneqq C_{\psi,t}\ket\psi$ is an arbitrary unnormalized state vector that can depend on $\ket{\psi}$ as well as on time.
The \ro can be written in its spectral decomposition
\begin{equation}
    \label{eq:RO_spectral}
    R_\psi = \sum_{i=1}^d\lambda_{i,\psi}\ketbra{\varphi_{i,\psi}},
\end{equation}
where $d$ is the dimension of the Hilbert space.
The stochastic part of the unraveling consists of jumps
\begin{equation}
    \label{eq:jump}
    \ket\psi\mapsto\ket{\varphi_{i,\psi}}
\end{equation}
to the eigenstate of $R_\psi$, happening with probability
\begin{equation}
    \label{eq:p_jump}
    p_{\psi\to\varphi_{i,\psi}} = \lambda_{i,\psi}\,dt,
\end{equation}
where $\lambda_{i,\psi}$ is the corresponding eigenvalue.
The deterministic evolution, on the other hand, reads
\begin{equation}
    \label{eq:det_evol}
    \begin{split}
        \ket{\psi(t+dt)} = \frac{(\id-iK_\psi\,dt)\ket\psi}{\norm{(\id-iK_\psi)\ket\psi}}
        = \ket\psi-iK_\psi\,dt\ket\psi+\frac{dt}2\tr R_\psi\,\ket\psi,
    \end{split}
\end{equation}
where
\begin{equation}
    \label{eq:K}
    K_\psi \coloneqq H-\frac i2\sum_\alpha\gamma_\alpha L_\alpha^\dagger L_\alpha-\frac i2 \ket{\Phi_\psi}\bra\psi
\end{equation}
is the non-linear effective Hamiltonian, depending as well on $\ket\psi$.
In \cite{Settimo2024}, it was shown that this unraveling technique can lead to positive rates whenever the P-divisibility condition \eqref{eq:P_div} holds.
Furthermore, the rates can remain positive even in some cases in which the dynamics is temporarily non-P-divisible.

If all rates $\lambda_{i,\psi}$ remain positive at all times for all trajectories, then the unraveling is said to be positive.
If, on the other hand, they turn temporarily negative, the unravelings can be equipped with reverse jumps \cite{Piilo-NMQJ-PRL, Piilo-NMQJ-PRA}, as it will be shown explicitly in Sec.~\ref{subsec:SSE_neg}.
Such reverse jumps are of the form
\begin{equation}
    \ket\psi = \ket{\varphi_{i,\psi^\prime}}\mapsto\ket{\psi^\prime},
\end{equation}
i.e. they are possible only if $\ket\psi$ is the target of a direct jump from some other state $\ket{\psi^\prime}$.
The probability of such a reverse jump is
\begin{equation}
    p_{\psi\to\psi^\prime}^{\text{rev}} = -\frac{p(\psi)}{p(\psi^\prime)}\lambda_{i,\psi^\prime}\,dt,
\end{equation}
where $\lambda_{i,\psi^\prime}<0$ is the eigenvalue of $R_{\psi^\prime}$ corresponding to the eigenstate $\ket\psi = \ket{\varphi_{i,\psi^\prime}}$.
Notice that the role of these reverse jumps is to revert a jump that would happen if the rate was positive.

% ------------------------------------ SSE ------------------------------------
\section{Stochastic \schro equation}
\label{sec:SSE}
The SSE corresponding to the \ro unraveling technique is now derived, first in the special case in which all rates $\lambda_{i,\psi}$ are positive and then in the general case of temporarily negative rates.

% ------------------------------------ Pos rates ------------------------------------
\subsection{Positive rates}
\label{subsec:SSE_pos}
The stochastic trajectories obey the non-linear SSE
\begin{equation}
    \label{eq:SSE}
    \ket{d\psi} = -i\tilde K_\psi\ket\psi\,dt + \sum_{i}\left(\ket{\varphi_{i,\psi}}-\ket\psi\right)\,dN_{i,\psi},
\end{equation}
where
\begin{equation}
    \tilde K_\psi \coloneqq K_\psi + \frac i2\tr [R_\psi]\,\id,
\end{equation}
and $dN_{i,\psi}$ are independent Poisson increments ($dN_{i,\psi}=0,1$) such that
\begin{equation}
    dN_{i,\psi}dN_{j,\psi} = \delta_{i,j}dN_{i,\psi},\qquad\mathbb E[dN_{i,\psi}] = \lambda_{i,\psi}\,dt,
\end{equation}
where $\mathbb E$ represents the expectation value with respect to the Poisson processes.

The SSE can be rewritten in terms of the projector $\ketbra\psi$ as
\begin{equation}
    \label{eq:dP}
    \begin{split}
        d(\ketbra\psi) =& \ket{d\psi}\bra\psi+\ket\psi\bra{d\psi}+\ketbra{d\psi}\\
        =&-i\left(\tilde K_\psi\ketbra\psi - \ketbra\psi\tilde K_\psi^\dagger\right)dt\\
        &+\sum_i\left(\ketbra{\varphi_{i,\psi}}-\ketbra\psi\right)dN_{i,\psi}.
    \end{split}
\end{equation}
Taking the expectation value, using Eq.~\eqref{eq:RO_spectral} and $\sum_i\lambda_{i,\psi}=\tr R_{\psi}$, then
\begin{equation}
    \label{eq:EdP}
    \begin{split}
        \mathbb E[d(\ketbra\psi)] =& -i\left(\tilde K_\psi\ketbra\psi - \ketbra\psi \tilde K_\psi^\dagger\right)dt + R_\psi\,dt - \tr [R_\psi]\ketbra\psi\,dt\\
        =& -i\left(K_\psi\ketbra\psi - \ketbra\psi K_\psi^\dagger\right)dt + R_\psi\,dt\\
        =&-i[H,\ketbra\psi]\,dt + \sum_\alpha\gamma_\alpha\left( L_\alpha\ketbra\psi L_\alpha^\dagger -\frac12\left\{L_\alpha^\dagger L_\alpha,\ketbra\psi\right\}\right)\,dt.
    \end{split}
\end{equation}
Employing Eq.~\eqref{eq:avg_rho} to write the state $\rho(t)$ and computing $d\rho = \int d\psi\,p(\psi)\,\mathbb E[d(\ketbra\psi)]$, it follows that
\begin{equation}
    \frac d{dt}\rho(t) = \mathcal L_t\left[\rho(t)\right]
\end{equation}
and indeed the average over all trajectories obeys the ME \eqref{eq:ME}.
Notice that when taking the expectation value $\mathbb E$, the non-linearity cancels out: Eqs.~\eqref{eq:SSE} and \eqref{eq:dP} are non-linear in both the jump and the driving term, since they both depend on $\ket{\Phi_\psi}$, while Eq.~\eqref{eq:EdP} is linear as expected.

The SSE \eqref{eq:SSE} is indeed equivalent to the unravelings, in the sense that it generates not only the same average dynamics but also the same trajectories.
Indeed, if all $dN_{i,\psi}=0$, then
\begin{equation}
    \ket{d\psi} = -iK_\psi\,dt\ket\psi+\frac{dt}2\tr R_\psi\,\ket\psi,
\end{equation}
which is the same of Eq.~\eqref{eq:det_evol}.
Notice that in this case, the deterministic evolution obeys the non-linear non-Hermitian norm-preserving \schro equation
\begin{equation}
    \frac d{dt}\ket{\psi(t)} = -i\tilde K_{\psi(t)}\ket{\psi(t)}.
\end{equation}
If, on the other hand, $dN_{i,\psi}=1$ (and thus $dN_{j,\psi}=0$ for all $j\ne i$), then a jump occurs
\begin{equation}
    \ket\psi\mapsto\ket\psi+\ket{d\psi} = \ket{\varphi_{i,\psi}} + O(dt),
\end{equation}
with the terms in $O(dt)$ that don't play any role, since the jumps happen with probability $\lambda_{i,\psi}\,dt$ and the terms in $O(dt^2)$ or higher are neglected.

% ------------------------------------ Neg rates ------------------------------------
\subsection{Negative rates}
\label{subsec:SSE_neg}
Let us now turn to the most general case in which $R_\psi$ has both positive and negative eigenvalues.
Without loss of generality, it is possible to separate the positive and negative part of the eigenvalues, and therefore of the \ro, as
\begin{equation}
    \lambda_{i,\psi}^\pm = \frac12(\abs{\lambda_{i,\psi}}\pm\lambda_{i,\psi})\ge0,\qquad
    R_\psi^\pm = \sum_i\lambda_{i,\psi}^\pm\ketbra{\varphi_{i,\psi}}\ge0,
\end{equation}
where $\lambda_i = \lambda_i^+-\lambda_i^-$ and the \ro can then be reconstructed as $R_\psi = R_\psi^+-R_\psi^-$.
The corresponding non-Markovian SSE will now have two types of independent Poisson processes $dN_{i,\psi}^\pm$, corresponding to the positive and negative eigenvalues of $R_\psi$ and it reads
\begin{equation}
    \label{eq:SSE_rev}
    \begin{split}
        \ket{d\psi} =&-i\tilde K_\psi\ket\psi\,dt + \sum_{i}\left(\ket{\varphi_{i,\psi}}-\ket\psi\right)\,dN_{i,\psi}^+\\
        &+\sum_i\int d\psi^\prime\,(\ket{\psi^\prime}-\ket\psi)\,dN_{i,\psi^\prime}^-.
    \end{split}
\end{equation}
The independence of the Poisson processes reads
\begin{gather}
    dN_{i,\psi}^+dN_{j,\psi}^+ = \delta_{i,j}\,dN_{i,\psi}^+,\qquad
    dN_{i,\psi}^+dN_{i,\psi^\prime}^-=0,\\
    dN_{i,\psi^\prime}^-dN_{j,\psi^{\prime\prime}}^- = \delta_{i,j}\,\delta\left(\ket{\psi^\prime}-\ket{\psi^{\prime\prime}}\right)\,dN_{i,\psi^\prime}^-,
\end{gather}
where $\delta\left(\ket{\psi^\prime}-\ket{\psi^{\prime\prime}}\right)$ is the Dirac delta on the set of pure states.
The expectation value of the positive Poisson process $dN^+_{i,\psi}$ is unchanged, i.e. $\mathbb E[dN_{i,\psi}^+] = \lambda_{i,\psi}^+\,dt$, while for the negative process it reads
\begin{equation}
    \label{eq:E_dN_-}
    \mathbb E\left[dN_{i,\psi^\prime}^-\right] = \frac{p(\psi)}{p(\psi^\prime)}\lambda_{i,\psi^\prime}^-\delta(\ket\psi-\ket{\varphi_{i,\psi^\prime}})\,dt.
\end{equation}
Note that the first line of Eq.~\eqref{eq:SSE_rev} is the same as Eq.~\eqref{eq:SSE} describing normal jumps, while the contribution of reverse jumps is given by the second line.
Furthermore, if $R_\psi\ge0$, then one has $\lambda^-_{i,\psi}=0$ and therefore the second line of Eq.~\eqref{eq:SSE_rev} vanishes.
Therefore, the SSE \eqref{eq:SSE} is a special case of the non-Markovian SSE \eqref{eq:SSE_rev}.

If, similarly to the positive rates case, one computes the expectation value of the increment $d(\ketbra\psi)$, then
\begin{equation}
    \label{eq:EdP_neg}
    \begin{split}
        \mathbb E[d(&\ketbra\psi)] = -i(\tilde K_\psi\ketbra\psi-\ketbra\psi\tilde K_\psi^\dagger)\,dt + R_\psi^+\,dt-\tr [R^+_\psi]\ketbra\psi\,dt\\
        &+\sum_i\int d\psi^\prime\,\frac{p(\psi^\prime)}{p(\psi)}\lambda_{i,\psi^\prime}^-\,\Big(\ketbra{\psi^\prime}-\ketbra\psi\Big)\,\delta\big(\ket\psi-\ket{\varphi_{i,\psi^\prime}}\big)\,dt.
    \end{split}
\end{equation}
The first line is the same of Eq.~\eqref{eq:EdP}, except for the fact that the second and the third term contain $R^+_\psi$ instead of $R_\psi$.

In order to explicitly evaluate the second line, one needs to consider the average with respect to $p(\psi)$.
This average is easily evaluated using the Dirac delta property $\int d\psi\,\delta\left(\ket\psi-\ket{\psi^\prime}\right)\,f[\psi] = f[\psi^\prime]$ for an arbitrary functional $f$, and it reads
\begin{equation}
    \label{eq:avg_rev}
    \begin{split}
        \int d\psi\, {p(\psi)}&\sum_i\int d\psi^\prime\,\frac{p(\psi^\prime)}{{p(\psi)}}\lambda_{i,\psi^\prime}^-\,\Big(\ketbra{\psi^\prime}-\ketbra\psi\Big)\,\delta\left(\ket\psi-\ket{\varphi_{i,\psi^\prime}}\right)\,dt\\
        =&\int d\psi^\prime\,p(\psi^\prime)\sum_i\lambda_{i,\psi^\prime}^-\,\Big(\ketbra{\psi^\prime}-\ketbra{\varphi_{i,\psi^\prime}}\Big)\,dt\\
        =&\int d\psi^\prime\,p(\psi^\prime)\Big(\tr [R_{\psi^\prime}^-]\ketbra{\psi^\prime}-R_{\psi^\prime}^-\Big)\,dt.
    \end{split}
\end{equation}
The average of the first line of Eq.~\eqref{eq:EdP_neg} is analogous to the average of Eq.~\eqref{eq:EdP} for positive rates.
Combining the two terms gives
\begin{equation}
    \int d\psi\,p(\psi)\,\mathbb E[d(\ketbra\psi)] = \mathcal L_t[\rho]\,dt
\end{equation}
which, in turn, implies that $\rho(t)$ obeys the ME \eqref{eq:ME} also in the presence of negative rates for the RO.

% ------------------------------------ Example ------------------------------------
\subsection{Breaking of positivity}
In \cite{Breuer2008}, it was shown that the stochastic unravelings arising from the Monte-Carlo Wave Function technique \cite{Dalibard-MCWF} equipped with reverse jumps fail whenever the dynamics violates positivity.
The same also holds for the SSE \eqref{eq:SSE_rev} and the \ro, independently of the particular transformation $\ket{\Phi_\psi}$ used in the definition of $R_\psi$.

Suppose that the ME \eqref{eq:ME} violates positivity at time $t_0$, then there must exist a solution $\rho(t)$ such that $\rho_0\coloneqq\rho(t_0)$ lies on the boundary of the set of quantum states, while $\rho(t_0+dt)$ lies outside it.
Let $\mu(t)$ be the minimum eigenvalue of $\rho(t)$ and $\ket{\xi(t)}$ the corresponding eigenvector.
Let $\rho_0 = \sum_j p_j\ketbra{\psi_j}$ be an ensemble representation of $\rho_0$ arising from the SSE \eqref{eq:SSE_rev}.
Since $\rho_0$ is on the boundary, then
\begin{equation}
    0=\mu(t_0) = \braket{\xi_0\vert\rho_0\vert\xi_0} = \sum_jp_j\abs{\braket{\xi_0\vert\psi_j}}^2,
\end{equation}
which implies that all $\ket{\psi_j}$ are orthogonal to $\ket{\xi_0} \coloneqq \ket{\xi(t_0)}$, i.e. $\braket{\xi_0\vert\psi_j}=0$.
Since $\rho(t_0+dt)$ lies outside the set of quantum states, then it must have a negative eigenvalue, which, in turn, implies
\begin{equation}
    \dot\mu(t_0) = \braket{\xi_0\vert\mathcal L_{t_0}[\rho_0]\vert\xi_0}
    =\sum_jp_j\braket{\xi_0\vert\mathcal J_{t_0}\left[\ketbra{\psi_j}\right]\vert\xi_0}<0.
\end{equation}
Since $\braket{\xi_0\vert\psi_j}=0$, adding $(\ket{\psi_j}\bra{\Phi_{\psi_j}} + \ket{\Phi_{\psi_j}}\bra{\psi_j})/2$ to $\mathcal J_{t_0}$ inside the mean value $\braket{\xi_0\vert\cdot\vert\xi_0}$ does not change $\dot\mu(t_0)$.
But, since this transformation is exactly the definition of the \ro \eqref{eq:RO}, this is equivalent to
\begin{equation}
    \label{eq:neg_mean_RO_pos_breaking}
    \sum_j p_j\braket{\xi_0\vert R_{\psi_j}\vert\xi_0} = \sum_j\sum_{i=1}^dp_j\lambda_{i,\psi_j}\abs{\braket{\xi_0\vert\varphi_{i,\psi_j}}}^2 <0.
\end{equation}
Therefore, there must be some $\lambda_{i,\psi_j}<0$ with $\braket{\xi_0\vert\varphi_{i,\psi_j}}\ne0$, so that the direct jump $\ket{\psi_j}\mapsto\ket{\varphi_{i,\psi_j}}$ has a non-zero component in the $\ket{\xi_0}$ direction.
This means that $\ket{\varphi_{i,\psi_j}}$ cannot be one of the $\ket{\psi_j}$ and therefore $p(\varphi_{i,\psi_j},t_0)=0$ and therefore the SSE breaks down.

Notice that Eq.~\eqref{eq:neg_mean_RO_pos_breaking} does not depend on the particular transformation $\ket{\Phi_{\psi_j}}$.
Therefore, the \ro unravelings lead to a singularity in the SSE \eqref{eq:SSE_rev} at the time in which positivity is violated independently of the particular way $R_\psi$ is chosen.
Therefore, the failure of the unravelings at time $t=t_0$ signal that the underlying dynamics is unphysical, and this happens independently of the arbitrary transformation $\ket{\Phi_{\psi_j}}$ considered.

% ------------------------------------ Example ------------------------------------
\section{Example}
\label{sec:example}
As an example of the use of the non-Markovian SSE \eqref{eq:SSE_rev}, a ME of the form
\begin{equation}
    \label{eq:ME_example}
    \mathcal L_t[\rho] = -i\beta[\sigma_z, \rho]+\sum_{\alpha=x,y,z}\gamma_\alpha\left(\sigma_\alpha\rho\sigma_\alpha^\dagger-\rho\right)
\end{equation}
is considered, where $\sigma_{x,y,z}$ are the Pauli matrices.
Without the driving, i.e. for $\beta = 0$, the dynamics is exactly solvable \cite{Chruscinski2022}, however, for $\beta\ne0$, such dynamics is highly non-trivial, especially for time-dependent driving.
The condition for P-divisibility \eqref{eq:P_div} can be rewritten as \cite{Chruscinski2022}
\begin{equation}
    \label{eq:P_div_example}
    \gamma_x+\gamma_y\ge0,\qquad\gamma_y+\gamma_z\ge0,\qquad\gamma_x+\gamma_z\ge0.
\end{equation}

The \ro formalism allows for very efficient simulations of the ME \eqref{eq:ME_example}.
Indeed, it is always possible to construct a transformation $\ket{\Phi_\psi}$ such that only three states are present in the average of Eq.~\eqref{eq:avg_rho}: the eigenstates $\ket0$, $\ket1$ of $\sigma_z$ and the deterministically evolved state $\ket{\psi_{\text{det}}(t)}$, i.e. the state evolved according to Eq.~\eqref{eq:det_evol}, conditioned to the no jumps happening.
The full form of $\ket{\Phi_\psi}$ can be derived in a simple and non-unique way and its form is too lengthy to be useful to report here.
The jump process only involves transitions of the form $\ket{\psi_{\text{det}}(t)}\mapsto\ket0,\ket1$ and $\ket0\leftrightarrow\ket1$.

The fact of having a small effective ensemble $\{\ket{\psi_{\text{det}}(t)}, \,\ket0,\,\ket1\}$ allows for high efficiency in the simulations, especially when reverse jumps are necessary, since one only needs to track the populations $p(\psi)$ of such three states, and therefore the Poisson processes $dN^-_{i,\psi}$ satisfying Eq.~\eqref{eq:E_dN_-} can be readily generated.

\begin{figure}
    \centering
    \includegraphics[width=.8\linewidth]{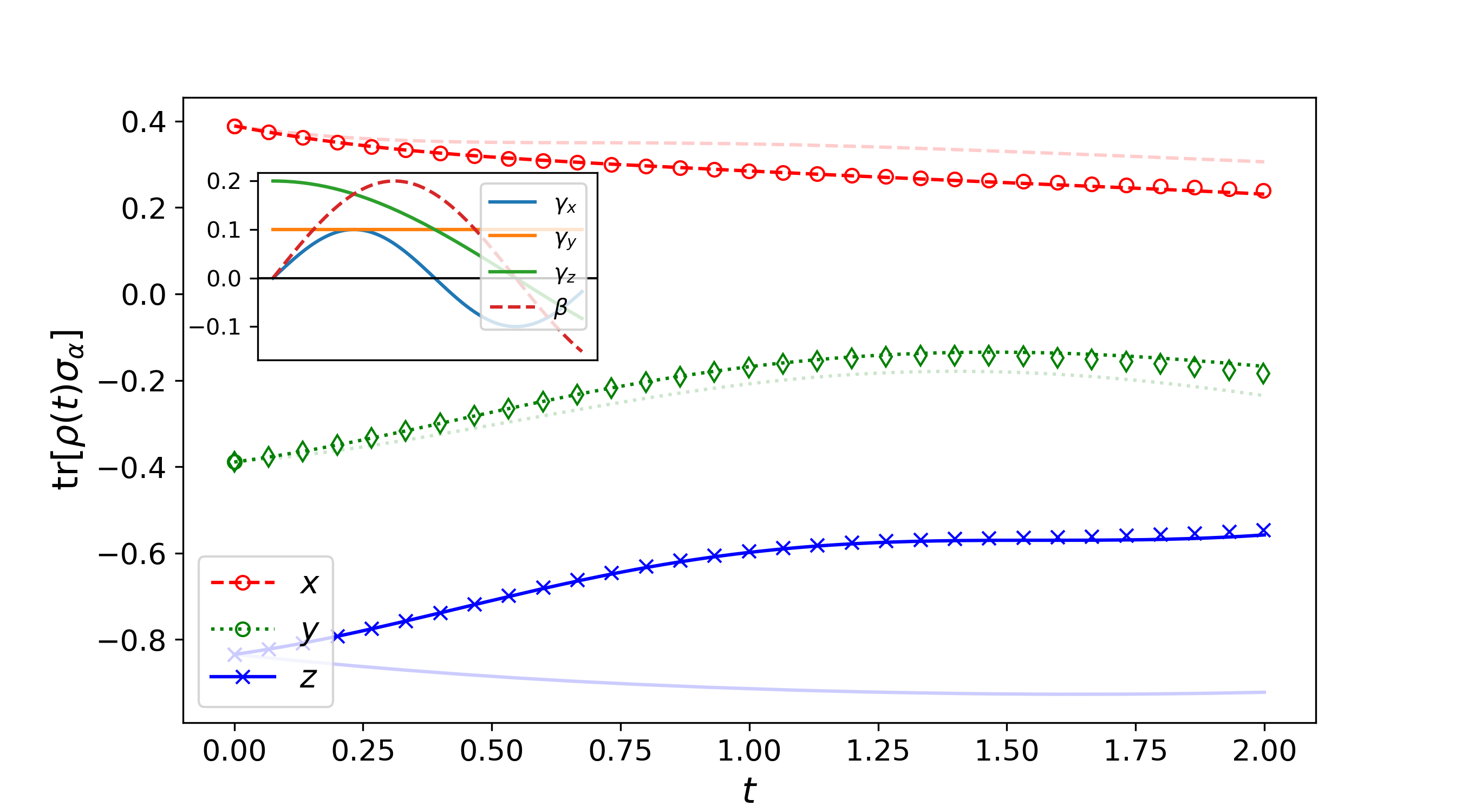}
    \caption{Unraveling of the ME \eqref{eq:ME_example} using the non-Markovian SSE \eqref{eq:SSE_rev}.
    The Bloch vector components are shown, and the unraveling matches the exact solution (dark lines) with small error.
    The lighter lines represent the Bloch vector components of $\ket{\psi_{\text{det}}(t)}$, i.e. the only time evolving state needed in the effective ensemble.
    Inset: rates $\gamma_\alpha$ and time-dependent driving strength $\beta$.}
    \label{fig:example}
\end{figure}

In Figure \ref{fig:example}, the solution of the ME \eqref{eq:ME_example} obtained by averaging of the non-Markovian SSE \eqref{eq:SSE_rev} is presented.
The rates $\gamma_\alpha$ are chosen in such a way that the condition \eqref{eq:P_div_example} is temporarily violated, and therefore the resulting dynamics is non-P-divisible and reverse jumps are indeed necessary in the unravelings.
Such rates are shown in the inset.
The \ro unraveling is not only efficient, but it also matches the exact solution (dark lines) with small error.
In lighter shades, the time evolution of $\ket{\psi_{\text{det}}(t)}$ is also shown.

% ------------------------------------ Conclusion ------------------------------------
\section{Conclusions}
\label{sec:conclusion}
In this work, a SSE for the \ro unraveling formalism was derived.
Such derivation allows for a proper formalization of the technique, both in the case of positive jump rates and in the presence of reverse jumps.
It was also shown that a failure of the SSE \eqref{eq:SSE_rev} implies a violation of positivity of the dynamical map $\Lambda_t$, thus providing a witness for unphysicality of the time evolution.
Noticeably, this condition does not depend on the non-linear transformation defining the \ro, and therefore can be readily checked.

The efficiency of this method, as well as the possibility of engineering the stochastic realizations, was exemplified by unraveling a non-P-divisible dynamics.
It was shown that it can be done with a small effective ensemble, which allows to easily compute the interdependence between stochastic trajectories when reverse jumps are considered.

\section*{Acknowledgements}
I thank Jyrki Piilo for the useful discussions and acknowledge financial support from Magnus Ehrnroothin S\"a\"ati\"o.

\printbibliography

\end{document}